\documentstyle[prb,aps,multicol,epsf]{revtex}

\begin{document}

\draft
\title{Kondo effect in XXZ spin chains}
\author{A. Furusaki$^*$}
\address{Yukawa Institute for Theoretical Physics, Kyoto University,
Kyoto 606-8502, Japan}
\author{T. Hikihara}
\address{Department of Physics, Kobe University, Rokkodai, Kobe 657,
Japan}
\date{\today}
\maketitle
\begin{abstract}
The Kondo effect in a one-dimensional spin-$\frac{1}{2}$ XXZ model
in the gapless XY regime ($-1<\Delta\le1$) is studied both analytically
and numerically. 
In our model an impurity spin ($S=1/2$) is coupled to a single spin
in the XXZ spin chain.
Perturbative renormalization-group (RG) analysis is performed for
various limiting cases to deduce low-energy fixed points.
It is shown that in the ground state the impurity spin is screened by
forming a singlet with a spin in the host XXZ chain.
In the antiferromagnetic side ($0<\Delta\le1$) the host chain is cut
into two semi-infinite chains by the singlet.
In the ferromagnetic side ($-1<\Delta<0$), on the other hand,
the host XXZ chain remains as a single chain through ^^ ^^ healing''
of a weakened bond in the low-energy (long-distance) limit.
The density matrix renormalization group method is used to study the
size scaling of finite-size energy gaps and the power-law decay of
correlation functions in the ground state.
The numerical results are in good agreement with the predictions of the
RG analysis.
Low-temperature behaviors of specific heat and susceptibility are also
discussed.
\end{abstract}

\begin{multicols}{2}

\section{Introduction}

There has been recent resurgence of interest in the Kondo
effect\cite{Kondo} in one-dimensional (1D) strongly-correlated
systems. 
In 1D interacting systems belonging to the universality class of the
Tomonaga-Luttinger (TL) liquids, a static impurity potential has a
drastic effect and is renormalized to infinity or zero, depending on
whether the interaction is repulsive or attractive.\cite{Kane} 
This anomalous response to a static impurity of TL liquids has
attracted a lot of attention and led to further studies on effects of
a dynamic impurity (typically a magnetic impurity) in a TL liquid.
A generalized Hubbard model with an impurity spin
($S=\frac{1}{2}$) and its variants have been studied by many authors.
It was found that the Kondo temperature, which is a typical energy
scale for host electrons to screen an impurity spin, has a power-law
dependence on the Kondo exchange
coupling.\cite{DHLee,FuruNaga,Schiller}
Properties of low-energy fixed points have been discussed using
perturbative renormalization group analysis\cite{FuruNaga} and the
boundary conformal field theory
approach.\cite{Frojdh,Durganandini,Granath} 
A recent Monte Carlo study\cite{Egger} on the susceptibility of an
impurity spin is consistent with anomalous power-law temperature
dependence conjectured earlier.\cite{FuruNaga} 
In addition to the models with a simple Kondo coupling, there are
some exactly solvable models in which an
impurity spin is coupled to the spin density of electrons via special
forms of the Kondo exchange coupling.\cite{Wang,Zvyagin}
The results obtained for these models using the Bethe-ansatz
technique, however, do not completely agree with the previous
studies,\cite{FuruNaga,Frojdh} and this remains as a question to be
resolved.

In this paper we consider a simplified model which we believe shares
common features with the above-mentioned Kondo effect in 1D
interacting electronic models like the Hubbard model.
We here focus on the spin sector and discard the charge degree of 
freedom.
This may correspond to the half-filled case in the original
electronic models.
The Hamiltonian of the system we discuss in this paper has the form
$H=H_0+H_K$, where $H_0$ describes the host $S=\frac{1}{2}$ XXZ spin
chain, 
\begin{equation}
H_0=J\sum_i\left(S^x_iS^x_{i+1}+S^y_iS^y_{i+1}
                 +\Delta S^z_iS^z_{i+1}\right),
\label{H_0}
\end{equation}
and $H_K$ the Kondo coupling,
\begin{equation}
H_K=J_K\left(S^x_0S^x_{\rm imp}+S^y_0S^y_{\rm imp}
          +\Delta S^z_0S^z_{\rm imp}\right).
\label{H_K}
\end{equation}
The size of the impurity spin is also assumed to be 1/2.
An important point of our model is absence of SU(2) spin rotation
symmetry. 
We assume $|\Delta|<1$ to ensure that the host XXZ spin chain has
gapless excitations.
For simplicity we have used the same parameter $\Delta$ in $H_0$ and
$H_K$.
We note that $\Delta$ in $H_0$ is an important parameter
controlling power-law behavior of various correlations while $\Delta$
in $H_K$ does not play any significant role in the following
discussions.
The Kondo coupling $J_K$ can be either antiferromagnetic or
ferromagnetic, but we will concentrate on the antiferromagnetic
case ($J_K>0$) in this paper.

Eggert and Affleck\cite{Eggert,Sorensen} studied, among various kinds of
disorder, the model at the isotropic point ($\Delta=1$).
They concluded that the impurity spin $\mbox{\boldmath$S$}_{\rm imp}$
forms a singlet with $\mbox{\boldmath$S$}_0$, and that the Heisenberg
chain is decoupled into two semi-infinite chains in the low-energy
limit. 
A leading irrelevant operator at the fixed point was identified and
shown to have scaling dimension 2.
It corresponds to exchange coupling between boundary spins of the
two decoupled chains. 
In this paper we extend their analysis to the XXZ case
($|\Delta|<1$).
We first bosonize the Hamiltonian and study its
renormalization group (RG) flows in the weak-coupling limit and in the 
strong-coupling limit.
We will argue that the system is renormalized to stable low-energy
fixed points where the impurity spin ($S=\frac{1}{2}$) is screened
exactly.
At the fixed points the boundary condition for the host XXZ spin chain
depends on the parameter $\Delta$ of the host chain:
For $0<\Delta\le1$ the spin chain is cut into two semi-infinite chains
with open boundary condition at $i=\pm1$.
On the other hand, for $-1<\Delta<0$ the host spin chain is not affected
much by the singlet and stays as a single chain.
Leading irrelevant operators at these fixed points have noninteger
scaling dimensions, yielding noninteger power-law temperature dependence
of impurity contribution to specific heat and susceptibility.
As evidences for this picture we will show finite-size scaling of
energy gap and spin-spin correlation functions in the ground state,
both of which are obtained by using the density matrix renormalization 
group (DMRG) method.
The numerical results are consistent with the picture drawn from the
perturbative RG analysis.
We note that our results are very different from a recent paper by
Liu,\cite{Liu} who studied the same model as ours using mysterious
transformations and calculated various quantities near a
strong-coupling fixed point.
For example, he obtained superlinear temperature dependence
($T^\alpha$: $\alpha>1$) for the impurity contribution to the specific
heat and vanishing susceptibility at zero temperature, both of which
cannot be correct from general grounds.

The plan of this paper is as follows.
In section II we discuss RG flows of our model using the
standard abelian bosonization method.
Impurity contributions to specific heat and susceptibility are also
discussed.
We show results of numerical DMRG calculations in section III and
compare them with conclusions of the perturbative RG in section II.
For simplicity we set $J=1$ throughout this paper.

\section{Perturbative renormalization group analysis}

\subsection{Weak-coupling limit}

We follow Ref.~\onlinecite{Eggert} and bosonize the Hamiltonian $H$.
Since the bosonization of the XXZ chain is a standard procedure,
we do not repeat the derivation of a bosonized Hamiltonian here.
After performing the Jordan-Wigner transformation and taking continuum 
limit, we find that $H_0$ reduces to a free-boson model,
\begin{equation}
H_0^{(b)}=
\frac{v}{2}\int dx\left[
\left(\frac{d\phi}{dx}\right)^2+\Pi^2
\right],
\label{bosonic-H_0}
\end{equation}
where $\Pi(x)$ is a conjugate operator to the bosonic field
$\phi(x)$: $[\phi(x),\Pi(y)]=i\delta(x-y)$.
The spin wave velocity $v$ is known to be
$v=(\pi/2\theta)\sin\theta$, where $\theta=\cos^{-1}\Delta$.
The spins in the chain can be represented in terms of
bosonic fields $\phi(x)$ and $\tilde\phi(x)$
($\Pi=d\tilde\phi/dx$):\cite{Eggert,Affleckpreprint}
\begin{mathletters}
\begin{eqnarray}
S^z_j&=&
\frac{1}{2\pi R}\frac{d\phi}{dx}
+c_1(-1)^j\sin\frac{\phi}{R},
\label{S^z}\\
S^+_j&=&
e^{i2\pi R\tilde\phi}\left[c_2\cos\frac{\phi}{R}+c_3(-1)^j\right],
\label{S^+}
\end{eqnarray}
\end{mathletters}\noindent
where $S^\pm_j=S^x_j\pm iS^y_j$.
Here $x\approx j$ and the $c_j$'s are numerical constants.
The lattice spacing is assumed to be unity.
The parameter $R$ in Eqs.\ (\ref{S^z}) and (\ref{S^+}) is related to
$\Delta$ in the original Hamiltonian (\ref{H_0}) as
\begin{equation}
R=
\left[
\frac{1}{2\pi}\left(1-\frac{1}{\pi}\cos^{-1}\Delta\right)
\right]^{1/2}.
\label{R}
\end{equation}
With the Gaussian form of $H_0^{(b)}$, we can immediately find the
scaling dimensions of operators $e^{ia\phi}$ and $e^{ia\tilde\phi}$,
both of which are $a^2/4\pi$.
Thus the dimensions of the staggered components of $S^z_i$ and $S^\pm$
are $(4\pi R^2)^{-1}$ and $\pi R^2$, respectively.

From Eqs.\ (\ref{S^z}) and (\ref{S^+}) the Kondo interaction term
$H_K$ becomes
\begin{eqnarray}
H_K^{(b)}&=&
S^+_{\rm imp}e^{-i2\pi R\tilde\phi(0)}
\left(\lambda_{F\perp}\cos\frac{\phi(0)}{R}+\lambda_{B\perp}\right)
+{\rm h.c.}
\nonumber\\
&&
+S^z_{\rm imp}
\left(\lambda_{Fz}\frac{d\phi(0)}{dx}
+\lambda_{Bz}\sin\frac{\phi(0)}{R}\right),
\label{bosonic-H_K}
\end{eqnarray}
where the couplings $\lambda$'s are proportional to $J_K$.
Since the impurity spin is coupled to a single spin
$\mbox{\boldmath$S$}_0$ in our model, we have backward Kondo
scattering terms proportional to
$\lambda_{Bz}$ and $\lambda_{B\perp}$.
These terms do not appear in some models where
$\mbox{\boldmath$S$}_{\rm imp}$ is coupled symmetrically to two
neighboring spins, say $\mbox{\boldmath$S$}_0$ and
$\mbox{\boldmath$S$}_1$.\cite{Andrei,KLee,Clarke}
These backscattering terms is an important ingredient of our model.
The backward spinflip scattering term ($\propto\lambda_{B\perp}$) has
scaling dimension $\pi R^2$ and is always a relevant operator.
This should be contrasted with the conventional Kondo problem in 3D,
where the Kondo interaction is a marginal operator of the form
$d\phi/dx$.
Therefore we conclude that the weak-coupling point ($J_K=0$) is
unstable for $-1<\Delta\le1$ independent of the sign of $J_K$, and the
system always flows to a strong-coupling regime.
This situation is quite similar to the Kondo effect in a TL
liquid.\cite{FuruNaga} 
To lowest order the scaling equation of the most divergent coupling
$\lambda_{B\perp}$ is given by
\begin{equation}
\frac{d\lambda_{B\perp}}{d\log L}=
(1-\pi R^2)\lambda_{B\perp},
\end{equation}
where $L$ is system size.
We thus expect that the energy scale $T_K$ at which the crossover from
weak coupling to strong coupling occurs should be
\begin{equation}
T_K\propto |\lambda_{B\perp}|^{1/(1-\pi R^2)}
\propto |J_K|^{1/(1-\pi R^2)}
\label{T_K}
\end{equation}
for $|J_K|\ll J (=1)$.
We identify this energy scale with the Kondo temperature.\cite{note1}

\subsection{Strong-coupling limit for $0<\Delta\le1$}
\begin{figure}
\narrowtext
\begin{center}
\leavevmode\epsfxsize=60mm
\epsfbox{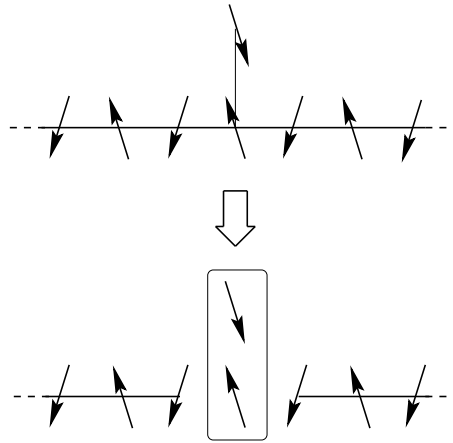}
\end{center}
\caption{
Schematic picture of renormalization to the strong-coupling fixed point
where the XXZ chain is cut by the singlet.
}
\label{fig:rgflow+}
\end{figure}

Let us consider the strong-coupling limit where $J_K\gg1$.
In this limit we first diagonalize $H_K$ and treat the coupling
between $\mbox{\boldmath$S$}_0$ and its neighbors
($\mbox{\boldmath$S$}_{\pm1}$) as weak perturbations.
The ground state of $H_K$ is a spin singlet
($\mbox{\boldmath$S$}_0+\mbox{\boldmath$S$}_{\rm imp}=0$).
In the limit $J_K\to\infty$ the system consists of the singlet and two
decoupled semi-infinite chains (SICs).
With very large but finite $J_K$, we derive effective interactions
acting on the subspace of the singlet plus the SICs using $1/J_K$
expansion.\cite{Nozieres}
Second order perturbation yields
\begin{eqnarray}
H_2&=&
-\frac{1}{2J_K(1+\Delta)}\left(S^+_1S^-_{-1}+S^-_1S^+_{-1}\right)
\nonumber\\
&&
-\frac{\Delta^2}{2J_K}S^z_1S^z_{-1}
+{\rm const}.
\label{second-order}
\end{eqnarray}
Higher order calculations also give the same form of interactions (and
irrelevant operators).
We now need to know the bosonization of these operators
$\mbox{\boldmath$S$}_{\pm1}$ at the boundaries of the SICs.
This was discussed in detail by Eggert and Affleck\cite{Eggert}
and we can simply borrow their results.
The open-boundary condition implies that the phase field $\phi(x)$
is fixed to be some constant at $x=0$.
To be specific, let us impose $\phi(0)=0$.
The left-going field $\phi_L(x)=\sqrt{\pi}[\phi(x)+\tilde\phi(x)]$ and
the right-going field $\phi_R(x)=\sqrt{\pi}[\phi(x)-\tilde\phi(x)]$
are no longer independent.
From these chiral fields we introduce two left-going fields:
\begin{mathletters}
\begin{eqnarray}
\phi_>(x)&=&\Theta(x)\phi_L(x)-\Theta(-x)\phi_R(-x),
\label{phi_>}\\
\phi_<(x)&=&\Theta(-x)\phi_L(x)-\Theta(x)\phi_R(-x),
\label{phi_<}
\end{eqnarray}
\end{mathletters}\noindent
where $\Theta(x)$ is a Heaviside step function.
The field $\phi_>(x)$ defined on ($-\infty,\infty$) describes bosonic
excitations in the SIC of the positive $x$ region
($\mbox{\boldmath$S$}_i$: $i>0$), and the other field $\phi_<$, also
defined on $(-\infty,\infty)$, describes excitations in the negative
$x$ region. 
Their commutation relations are
$[\phi_>(x),\phi_>(y)]=[\phi_<(x),\phi_<(y)]=-i\pi{\rm sgn}(x-y)$
and $[\phi_>(x),\phi_<(y)]=0$.
Their dynamics is governed by the Hamiltonian
\begin{equation}
H_{\rm SIC}=\frac{v}{4\pi}\int dx
\left[
\left(\frac{d\phi_>}{dx}\right)^2+\left(\frac{d\phi_<}{dx}\right)^2
\right].
\label{H_SIC}
\end{equation}
With these fields the boundary spins can be written as
\begin{mathletters}
\begin{eqnarray}
S^\pm_1&\propto& \exp\left[\pm i2\sqrt{\pi}R\phi_>(0)\right],\quad
S^z_1\propto\frac{d\phi_>(0)}{dx},
\label{S_1}
\\
S^\pm_{-1}&\propto&\exp\left[\pm i2\sqrt{\pi}R\phi_<(0)\right],\quad
S^z_{-1}\propto\frac{d\phi_<(0)}{dx}.
\label{S_-1}
\end{eqnarray}
\end{mathletters}\noindent
The scaling dimension of $S^z_{\pm 1}$ is 1 and that of
$S^\pm_{\pm 1}$ is $2\pi R^2$.
In general, the vertex operators $e^{ia\phi}$ and $e^{ia\tilde\phi}$
have dimension $a^2/2$ at boundaries.
We thus find that, among possible interactions generated by the
$1/J_K$ expansions, $S^+_1S^-_{-1}+S^-_1S^+_{-1}$ is most dangerous
and has dimension $4\pi R^2$.
This operator is irrelevant when $0<\Delta\le1$.
Therefore we may conclude that, when the anisotropy parameter $\Delta$
of the host XXZ spin chain is $0<\Delta\le1$, the infrared stable
fixed point corresponds to the limit $J_K\to\infty$, where the system
is decoupled into a singlet and two semi-infinite XXZ spin chains;
see Fig.~\ref{fig:rgflow+}.
The singlet acts like an infinitely high potential barrier for
excitations in the spin chain and effectively cuts it into two SICs.
If the host spin chain is of finite length containing $L$ spins and if
we apply the periodic boundary condition, then its low-energy fixed
point is an open spin chain consisting of $L-1$ spins, in addition to
a decoupled spin singlet formed from the impurity spin and a spin
originally in the host spin chain.\cite{reality} 
This strong-coupling fixed point is very similar to the one found for
the Kondo effect in electronic TL liquids.\cite{FuruNaga}

\noindent

The above result is a natural generalization of the conclusion of Eggert
and Affleck to the case $0<\Delta<1$.
In their case the low-energy fixed point is a singlet plus decoupled
two semi-infinite Heisenberg spin chains, and the leading irrelevant
operator at the fixed point is a dimension 2 operator,
$\mbox{\boldmath$S$}_1\cdot\mbox{\boldmath$S$}_{-1}$.
In our case the operator $S^+_1S^-_{-1}+S^-_1S^+_{-1}$ has a smaller
scaling dimension than $S^z_1S^z_{-1}$ because of the absence of the
SU(2) symmetry.
Since its dimension $4\pi R^2$ is in general noninteger, we may expect
that it should give anomalous power-law temperature dependence to
various quantities.

The coupling to the Kondo impurity gives rise to an extra contribution
to the specific heat and the spin susceptibility, which we denote
$\delta C$ and $\delta\chi$.
Their temperature dependence near the strong-coupling fixed point is
determined by the leading irrelevant operator,
$\widehat{O}_1=g_1\left(S^+_1S^-_{-1}+S^-_1S^+_{-1}\right)$, where
$g_1$ is a coupling constant.
To obtain leading temperature dependence we may use a perturbation
expansion in $\widehat{O}_1$.\cite{Affleck}

We first estimate $\delta C$.
Up to second order, the change in the free energy is given by
\begin{eqnarray}
\delta F&=&
-\int^{\beta/2}_0d\tau\langle\widehat{O}_1(\tau)\widehat{O}_1(0)\rangle
\nonumber\\
&\propto&
-\int^{\beta/2}_{\tau_c}d\tau
\left(\frac{\pi T\tau_c}{\sin\pi T\tau}\right)^{2d},
\label{delta_F}
\end{eqnarray}
where $d=4\pi R^2$, $\beta$ is inverse of the temperature $T$,
$\widehat{O}_1(\tau)\equiv e^{\tau H_{\rm SIC}}
\widehat{O}e^{-\tau H_{\rm SIC}}$, and $\tau_c$ is a cutoff to
regularize the integral.
Note that there is no first-order contribution of $\widehat{O}_1$
to $\delta F$.
The low-temperature expansion of the integral in Eq.\ (\ref{delta_F})
for general $d$ reads
\begin{eqnarray}
&&
\int^{\beta/2}_{\tau_c}
\left(\frac{\pi T\tau_c}{\sin\pi T\tau}\right)^{2d}d\tau
-\frac{\tau_c}{2d-1}
\nonumber\\
&&
=
\cases{
-\frac{\tau_c}{3}(\pi T\tau_c)^2,
&$d=1$,\cr
\frac{\tau_c(d-1)}{2d-1}(\pi T\tau_c)^{2d-1}
B(\textstyle{\frac{1}{2},\frac{3}{2}-d}),
&$1<d<\frac{3}{2}$,\cr
\frac{\tau_c}{2}(\pi T\tau_c)^2\log(1/\pi T\tau_c),
&$d=\frac{3}{2}$,\cr
\frac{\tau_cd(\pi T\tau_c)^2}{3(2d-3)},
&$\frac{3}{2}<d<\frac{5}{2}$,\cr}
\label{low-T-expansion}
\end{eqnarray}
where $B(a,b)$ is the beta function.
Note that any irrelevant operator with dimension $d>3/2$ generates a
positive $T^2$ term.
From these equations we get
\begin{equation}
\delta C\propto
\cases{
(d-1)^2T^{2d-2}/(3-2d),&$1<d<\frac{3}{2}$,\cr
T\log(1/\pi T\tau_c),&$d=\frac{3}{2}$,\cr
T/(2d-3),&$\frac{3}{2}<d<\frac{5}{2}$,\cr}
\label{delta-C}
\end{equation}
in the low-temperature limit.
Since $d=4\pi R^2$, the boundary case $d=\frac{3}{2}$ corresponds to
$\Delta=1/\sqrt{2}$.
When $0<\Delta<1/\sqrt{2}$, $\delta C$ is proportional to $T^{8\pi
R^2-2}$ with the exponent changing from 0 to 1 as $\Delta$ varying from 0
to $1/\sqrt{2}$.
This anomalous power-law behavior is reminiscent of the Kondo effect in TL
liquids.\cite{FuruNaga}
The log correction appears at $\Delta=1/\sqrt{2}$ when the dimension of
the leading irrelevant operator becomes $3/2$.
This is mathematically the same as in the two-channel Kondo
problem.\cite{Affleck} 
When $1/\sqrt{2}<\Delta\le1$, the leading term of $\delta C$ is
proportional to $T$.

We next consider $\delta \chi$.
Here we need to distinguish two kinds of spin susceptibilities:
one responding to a magnetic field applied in the $z$ direction and the
other responding to the one in the $xy$ plane. 
We shall call them $\delta\chi_z$ and $\delta\chi_\perp$,
respectively. 
Suppose we apply a magnetic field locally\cite{locally} only to
$\mbox{\boldmath$S$}_{\rm imp}$ such that the perturbation, 
\begin{equation}
H_h=h_zS^z_{\rm imp} + h_xS^x_{\rm imp},
\label{H_h}
\end{equation}
is added to the Hamiltonian.
Using the $1/J_K$ expansion again, we can generate effective
interactions induced by $H_h$ in the Hilbert space of the singlet plus
the SICs (Fig.~\ref{fig:rgflow+}).
From the symmetry we expect to have the following operators in
addition to other less relevant ones:
$\widehat{O}_{h1}=h_z(S^z_1+S^z_{-1})$,
$\widehat{O}_{h2}=h^2_z(S^+_1S^-_{-1}+S^-_1S^+_{-1})$, and
$\widehat{O}_{h3}=h_x(S^x_1+S^x_{-1})$.
In terms of the bosonic fields they may be written as
$\widehat{O}_{h1}\propto
h_z[\partial_x\phi_>(0)+\partial_x\phi_<(0)]$, 
$\widehat{O}_{h2}\propto
  h^2_z\cos\{2\sqrt{\pi}R[\phi_>(0)-\phi_<(0)]\}$,
and
$\widehat{O}_{h3}\propto h_x\{\cos[2\sqrt{\pi}R\phi_>(0)]
   +\cos[2\sqrt{\pi}R\phi_<(0)]\}$,
whose scaling dimensions are 1, $4\pi R^2$, and $2\pi R^2$.
We can now estimate $\delta F$ induced by these operators using
Eq.~(\ref{low-T-expansion}) and obtain
$\delta\chi_\alpha=-\partial^2\delta F/\partial h^2_\alpha|_{h=0}$.
One point to be mentioned is that products of $\widehat{O}_1$ and
$\widehat{O}_{h2}$ can contribute a term $h^2_zT^{8\pi R^2-1}$ to
$\delta F$, leading to a term proportional to $T^{8\pi R^2-1}$ in
$\delta\chi_z$. 
From these considerations, we conclude that in the low-temperature
limit $\delta\chi$ has the following form:
\begin{mathletters}
\begin{eqnarray}
\delta\chi_z(T)-\delta\chi_z(0)&\propto&
\cases{
\frac{4\pi R^2-1}{8\pi R^2-3}T^{8\pi R^2-1},
&$0<\Delta<1/\sqrt{2}$,\cr
T^2\log(1/T)
&$\Delta=1/\sqrt{2}$,\cr
T^2
&$1/\sqrt{2}<\Delta\le1$,\cr
}
\label{delta-chi_z}
\\
\delta\chi_\perp(T)-\delta\chi_\perp(0)&\propto&
T^{4\pi R^2-1},
\quad 0<\Delta\le1.
\label{delta-chi_perp}
\end{eqnarray}
\end{mathletters}\noindent
We note that there is always a contribution proportional to $T^2$
coming from irrelevant operators.
When $0<\Delta\ll1$, the term $T^{8\pi R^2-1}$ might be difficult to
observe, because of its small coefficient $(\propto 4\pi R^2-1)$,
compared with the $T^2$ term.
We also note that in general the zero-temperature limit of the
susceptibility $\delta\chi(0)$ is of order $1/T_K$.

\subsection{Strong-coupling limit for $-1<\Delta<0$}

When the parameter $\Delta$ in the host spin chain is in the range
$-1<\Delta<0$, the dimension of the operator
$S^+_1S^-_{-1}+S^-_1S^+_{-1}$ is smaller than 1 and is relevant.
This means that the open-boundary fixed point discussed in the
previous subsection cannot be a low-energy fixed point when
$-1<\Delta<0$.
Both limits $J_K\to0$ and $J_K\to\infty$ in the original Hamiltonian
$H_0+H_K$ are unstable.
We thus need to find a nontrivial fixed point.

Let us for the moment forget the singlet of
$\mbox{\boldmath$S$}_{\rm imp}$ and $\mbox{\boldmath$S$}_0$, and
concentrate on the rest of the spins.
That is, we consider the two semi-infinite spin chains weakly coupled
by a ferromagnetic exchange interaction $H_2$:
\begin{eqnarray}
H_\lambda&=&
\sum^\infty_{i=1}
\left(S^x_iS^x_{i+1}+S^y_iS^y_{i+1}+\Delta S^z_iS^z_{i+1}\right)
\nonumber\\
&&+
\sum^\infty_{i=1}
\left(
S^x_{-i}S^x_{-i-1}+S^y_{-i}S^y_{-i-1}+\Delta S^z_{-i}S^z_{-i-1}
\right)
\nonumber\\
&&
-\lambda\left(S^x_1S^x_{-1}+S^y_1S^y_{-1}\right),
\label{H_lambda}
\end{eqnarray}
where $0<\lambda\ll1$.
We have dropped the irrelevant $S^z_1S^z_{-1}$ term.
Now we rotate $\mbox{\boldmath$S$}_i$ ($i>0$) around the $z$ axis by
$\pi$, which changes the sign of $\lambda$ ($-\lambda\to\lambda$) in
$H_\lambda$.
We then apply RG transformation.
Since $S^x_1S^x_{-1}+S^y_1S^y_{-1}$ is relevant, the coupling
$\lambda$ grows as the energy scale decreases.
The $S^z_1S^z_{-1}$ term is also generated in the course of the RG
transformation.
Thus, the two chains get coupled stronger at lower energy scale.
We next consider the opposite limit where the two chains are well
connected but one bond is slightly disturbed. 
This is described by the Hamiltonian,
\begin{eqnarray}
H_\varepsilon&=&
\sum^\infty_{i=1}
\left(S^x_iS^x_{i+1}+S^y_iS^y_{i+1}+\Delta S^z_iS^z_{i+1}\right)
\nonumber\\
&&+
\sum^\infty_{i=1}
\left(
S^x_{-i}S^x_{-i-1}+S^y_{-i}S^y_{-i-1}+\Delta S^z_{-i}S^z_{-i-1}
\right)
\nonumber\\
&&
+(1-\varepsilon_\perp)\left(S^x_1S^x_{-1}+S^y_1S^y_{-1}\right)
+(1-\varepsilon_z)\Delta S^z_1S^z_{-1},
\label{H_varepsilon}
\end{eqnarray}
where $0<\varepsilon_\perp,\varepsilon_z\ll1$.
Bosonizing this Hamiltonian as in Sec.~IIA, we find that the
perturbations ($\propto\varepsilon$) give the spin-Peierlse operator
$\sin[\phi(0)/R]$ of dimension $(4\pi R^2)^{-1}$ and dimension 2
operators like $(\partial\phi/\partial x)^2$.
Since they are irrelevant ($\varepsilon_\perp,\varepsilon_z\to0$ in
the low-energy limit), we recover a pure XXZ spin chain.
It is tempting to assume that the RG trajectories starting from the
unstable point describing two weakly coupled chains
[Eq.~(\ref{H_lambda})]  continuously flow
to the stable fixed point of the pure XXZ chain.
Although we cannot prove it, we believe this is what actually happens.
We note that this phenomenon is closely related to the well-known result
that a backwardscattering potential is renormalized to zero for fermions
interacting with mutual attractive interactions.\cite{Kane}
It is also similar to the ^^ ^^ healing'' of weak bonds which
Eggert and Affleck found for the isotropic Heisenberg chain with two
symmetrically perturbed bonds.\cite{Eggert}
Coming back to the Hamiltonian $H_\lambda$, we conclude that its
low-energy fixed point is a pure  XXZ spin chain with the spins
$\mbox{\boldmath$S$}_i$ ($i>0$) rotated around the $z$ axis by $\pi$.

\begin{figure}
\narrowtext
\begin{center}
\leavevmode\epsfxsize=60mm
\epsfbox{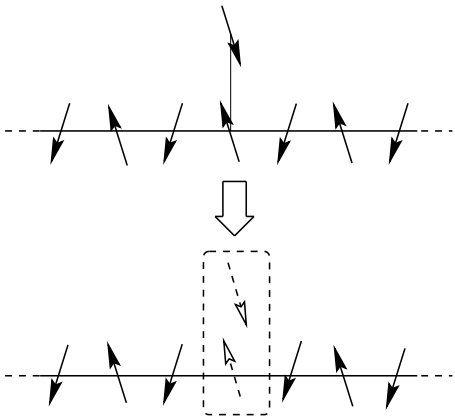}
\end{center}
\caption{
Schematic picture of renormalization to the stable fixed point
consisting of a singlet weakly coupled to a spin chain.
The singlet looks ^^ ^^ transparent'' for low-energy excitations
in the chain. 
}
\label{fig:rgflow-}
\end{figure}\noindent

\begin{figure}
\narrowtext
\begin{center}
\leavevmode\epsfxsize=60mm
\epsfbox{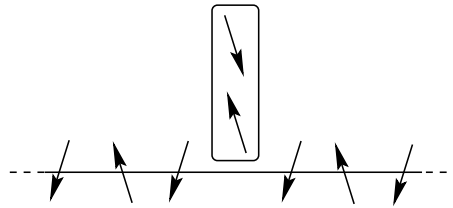}
\end{center}
\caption{
Schematic picture of the low-energy fixed point.
}
\label{fig:rgflow-2}
\end{figure}

We now return to our Kondo problem.
What we have found so far is that (i) the Kondo coupling is a relevant
operator at the weak-coupling point and leads to a singlet formation
and that (ii) weakly coupled spin chains are renormalized to a
strongly coupled single chain.
Combining these two observations together, we propose the model
schematically shown in Fig.~\ref{fig:rgflow-} as a candidate for the
low-energy fixed point.
The model consists of the singlet of $\mbox{\boldmath$S$}_{\rm imp}$
and $\mbox{\boldmath$S$}_0$ on top of the pure XXZ chain where spins
are rotated as discussed in the last paragraph.
An important point is that low-energy excitations are  spin density
fluctuations of long wave length in the chain and that for these
low-energy excitations the singlet has essentially no effect.
In other words, the singlet is ^^ ^^ transparent'' for them.
At short-length scale there is a coupling between
$\mbox{\boldmath$S$}_0$ and its neighbors
($\mbox{\boldmath$S$}_1+\mbox{\boldmath$S$}_{-1}$).
We assume that, as far as low-energy physics is concerned, the singlet
is rigid and can be broken only virtually by the weak coupling of
$\mbox{\boldmath$S$}_0$ to the spin chain.
Thus, the stable fixed point may also be represented schematically as
in Fig.~\ref{fig:rgflow-2}.
From the assumption of the rigid singlet, we can integrate out it to
get effective interactions
$\widehat{O}_2=S^x_1S^x_{-1}+S^y_1S^y_{-1}$ and
$\widehat{O}_3=S^z_1S^z_{-1}$ for the low-energy excitations in
the spin chain.
In the boson representation they are linear combinations of
$\sin(\phi/R)$, $(\partial\phi/\partial x)^2$, and
$(\partial\phi/\partial t)^2$, 
which are irrelevant operators for the spin chain with the parameter
$\Delta$ in the range $-1<\Delta<0$.
Hence the model is stable against weak perturbations,
and we conjecture that the above model gives a correct picture of the
strong-coupling fixed point for the case $-1<\Delta<0$.
Although it is impossible to show analytically that the RG trajectories
leaving from the unstable weak-coupling point reach this fixed point
(Figs.~\ref{fig:rgflow-} and \ref{fig:rgflow-2}), the numerical results
we show in the next section provide good evidences for our picture.

\noindent

Assuming that our Kondo model is indeed renormalized to the
strong-coupling fixed point of Fig.~\ref{fig:rgflow-}, we can obtain
leading temperature dependences of $\delta C$ and $\delta\chi$ as in the
last subsection.
Since we know that a leading irrelevant operator at the fixed point is
among the operators $\sin[\phi(0)/R]$, $[\partial\phi(0)/\partial x]^2$,
and $[\partial\phi(0)/\partial t]^2$, we find that the low-temperature
behavior of $\delta C$ is given by Eq.~(\ref{delta-C}) with
$d=(4\pi R^2)^{-1}$.
We thus get
\begin{equation}
\delta C\propto\cases{
T,&$-1<\Delta<-1/2$,\cr
T\log(1/T),&$\Delta=-1/2$,\cr
T^{1/(2\pi R^2)-2},&$-1/2<\Delta<0$.}
\end{equation}

When a weak magnetic field is applied to $\mbox{\boldmath$S$}_{\rm imp}$,
we obtain the operators $\widehat{O}_{h1}=h_z(S^z_1+S^z_{-1})$,
$\widehat{O}_{h2}=h^2_z(S^+_1S^-_{-1}+S^-_1S^+_{-1})$, and
$\widehat{O}_{h3}=h_x(S^x_1+S^x_{-1})$ after integrating out the singlet.
Since these operators are not boundary operators at the fixed point of
our interest, the scaling dimensions of $\widehat{O}_{h2}$ and
$\widehat{O}_{h3}$ are different from the open-boundary case.
Here we use the bosonization formulas (\ref{S^z}) and (\ref{S^+}) and
find that the dimensions of $\widehat{O}_{h2}$ and $\widehat{O}_{h3}$
are $(4\pi R^2)^{-1}$ and $\pi R^2$, respectively.
We then obtain the following low-temperature behavior:
\begin{mathletters}
\begin{eqnarray}
&&
\delta\chi_z(T)-\delta\chi_z(0)
\nonumber\\
&&\qquad
\propto
\cases{
\frac{1-4\pi R^2}{6\pi R^2-1}T^{(1/2\pi R^2)-1},
&$-\frac{1}{2}<\Delta<0$,\cr
T^2\log(1/T),
&$\Delta=-\frac{1}{2}$,\cr
T^2/(1-6\pi R^2),
&$-1<\Delta<-\frac{1}{2}$,\cr
}
\label{delta-chi_z-2}
\\
&&
\delta\chi_\perp\propto T^{2\pi R^2-1}.
\label{delta-chi_perp-2}
\end{eqnarray}
\end{mathletters}\noindent

\subsection{Strong-coupling limit of the XY case ($\Delta=0$)}

We briefly comment on the low-energy fixed point for the XY case.
Since this is exactly on the border of the two cases discussed in
Secs.~IIB and IIC, we naturally expect that a picture for the fixed
point of the $\Delta=0$ case should be something in between
Figs.~\ref{fig:rgflow+} and \ref{fig:rgflow-}.
That is, the singlet of $\mbox{\boldmath$S$}_{\rm imp}$ and
$\mbox{\boldmath$S$}_0$ does not completely cut the host XXZ spin
chain into two pieces.
The weakened connection between $\mbox{\boldmath$S$}_1$ and
$\mbox{\boldmath$S$}_{-1}$ is not healed as in the negative $\Delta$
case.
This is because at the open-boundary fixed point $(J_K=\infty)$
the operator $S^+_1S^-_{-1}+S^-_1S^+_{-1}$ is a marginal operator.
We expect that the impurity contribution to the specific heat and
the susceptibilities have the following low-temperature limit:
\begin{mathletters}
\begin{eqnarray}
&&\delta C\propto T,
\label{delta-C-3}
\\
&&
\delta\chi_z(T)-\delta\chi_z(0)\propto T^2,
\label{delta-chi_z-3}
\\
&&
\delta\chi_\perp(T)\propto\log(1/T).
\label{delta-chi_perp-3}
\end{eqnarray}
\end{mathletters}\noindent

\section{Results of DMRG calculations}

\subsection{Numerical methods}

In this section we present our numerical results.
The Hamiltonian we studied is $H_0+H_K$, Eqs.\ (\ref{H_0}) and
(\ref{H_K}).
The site index $i$ in Eq.\ (\ref{H_0}) runs from $-l$ to $l-1$, and
the total number of spins in the host XXZ chain is $L=2l+1$.
We impose the open boundary condition at the left and right ends of the
host XXZ chain.
Using the DMRG method proposed by White,\cite{White1} we have
calculated lowest energy gap and spin correlation functions in the
ground state.
In order to accelerate the numerical calculation, we have employed the
improved algorithm proposed by White.\cite{White2}
We have also used the finite system method to achieve high accuracy.
Up to 100 states were kept for each block and the truncation 
error is typically $10^{-8}$.
This error is directly related to the accuracy of energy.

\subsection{Numerical results for $\Delta=0.5$}

\begin{figure}
\narrowtext
\begin{center}
\leavevmode\epsfxsize=80mm
\epsfbox{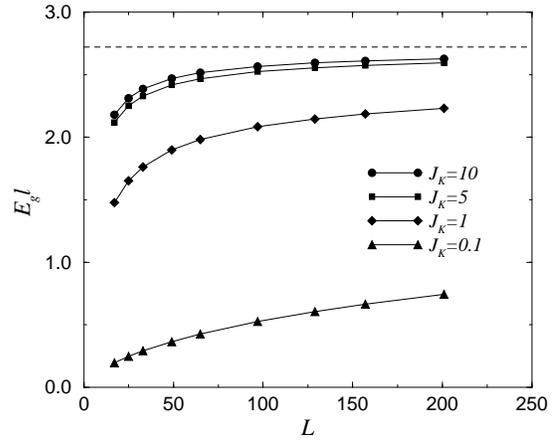}
\end{center}
\caption{
Energy gap $E_g$ as a function of system size $L$ for $\Delta=0.5$.
The data points are the gap computed for $L=17$, 25, 33, 49, 65, 94,
129, 157, and 201 [$L=1$ (mod 4)].
The dashed line represents the infinite-$L$ limit,
$E_gl=\protect{\sqrt{3}}\pi/2$.}
\label{fig:gap_d.5}
\end{figure}

As a typical case of $0<\Delta<1$ we have chosen $\Delta=0.5$.
In this case $R=1/\sqrt{3\pi}$ and $v=3\sqrt{3}/4$.
With this choice we have computed lowest energy gap $E_g$ for chains
of $L=1$ (mod 4).
Numerical results of the finite-size gap is shown in
Fig.~\ref{fig:gap_d.5}. 
The energy gap is difference between the lowest energy in the sector
$S^z_{\rm tot}=0$ and that in the sector $S^z_{\rm tot}=1$.
According to the RG analysis in Sec.~IIB, the ground state of
a sufficiently long chain is described as two decoupled chains, each
having $(L-1)/2$ spins, plus a rigid spin singlet of
$\mbox{\boldmath$S$}_0$ and $\mbox{\boldmath$S$}_{\rm imp}$ in between
them. 
Note that $(L-1)/2=l$ is an even integer.

\noindent

\begin{figure}
\narrowtext
\begin{center}
\leavevmode\epsfxsize=83mm
\epsfbox{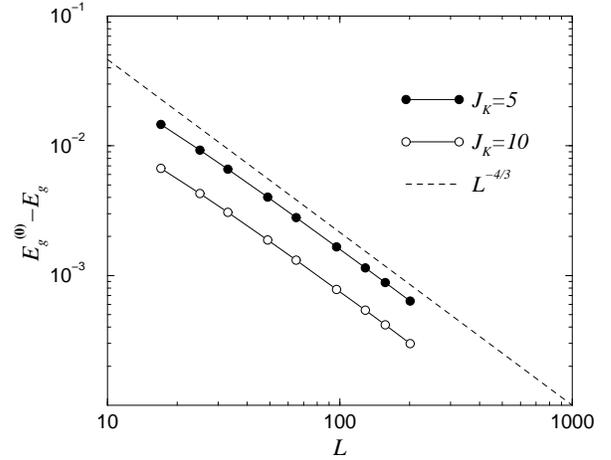}
\end{center}
\caption{
Size dependence of the correction to the energy gap $E^{(0)}_g$, which is
the gap calculated in the limit $J_K=\infty$.
The dashed line represents the theoretically predicted $L^{-4/3}$
dependence.}
\label{fig:diff_gap_d.5}
\end{figure}

To interpret finite-size scaling of the data, let us bosonize the two
open XXZ chains of length $l$, following
Refs.~\onlinecite{Eggert} and \onlinecite{Qin}.
The mode expansions of the phase fields are given by
\begin{eqnarray}
\phi_\mu(x,t)&=&
\pi R+\widehat{Q}_\mu\frac{x}{l}
\nonumber\\
&&
+\sum_{n>0}\frac{\sin k_nx}{\sqrt{\pi n}}
\left(a_{n\mu}e^{-ik_nvt}+a_{n\mu}^\dagger e^{ik_nvt}\right),
\label{phi-mode}
\\
\tilde\phi_\mu(x,t)&=&
\tilde\phi_{0\mu}+\widehat{Q}_\mu\frac{vt}{l}
\nonumber\\
&&
+i\sum_{n>0}\frac{\cos k_nx}{\sqrt{\pi n}}
\left(a_{n\mu}e^{-ik_nvt}-a_{n\mu}^\dagger e^{ik_nvt}\right),
\label{tilde-phi-mode}
\end{eqnarray}
where $k_n=\pi n/l$ and the operators obey the commutation relations
$[\tilde\phi_{0\mu},\widehat{Q}_\nu]=i\delta_{\mu,\nu}$ and
$[a_{m\mu},a^\dagger_{n\nu}]=\delta_{m,n}\delta_{\mu,\nu}$
$(\mu,\nu=l$ or $r)$.
The suffix $l$ and $r$ stand for the left ($\mbox{\boldmath$S$}_i$:
$i<0$) and right ($\mbox{\boldmath$S$}_i$: $i>0$) spin chain,
respectively.
The fields $\phi_l$ and $\tilde\phi_l$ ($\phi_r$ and $\tilde\phi_r$)
are therefore defined in the negative (positive) $x$ region, and
$\phi_l+\phi_r$ and $\tilde\phi_l+\tilde\phi_r$ correspond to $\phi$
and $\tilde\phi$ in Eq.\ (\ref{bosonic-H_0}).
Note that $\phi_l$ and $\phi_r$ are different from $\phi_<$ and
$\phi_>$.
Substituting Eqs.\ (\ref{phi-mode}) and (\ref{tilde-phi-mode}) into
Eq.\ (\ref{bosonic-H_0}) yields the Hamiltonian of the $\mu$ chain
\begin{equation}
H_\mu=
\frac{\pi v}{l}\left(
\frac{\widehat{Q}^2_\mu}{2\pi}
+\sum_{n>0}na^\dagger_{n\mu}a_{n\mu}-\frac{1}{24}
\right).
\label{H_mu}
\end{equation}
Its energy eigenvalue and eigen functions are
\begin{eqnarray}
&&E_{\mu}=
\frac{\pi v}{l}\left[
2\pi R^2\left(S^z_\mu\right)^2+\sum_{n>0}nm_{n\mu}-\frac{1}{24}
\right],
\label{E_mu}
\\
&&
|S^z_\mu,\{m_{n\mu}\}\rangle=
\exp\left(i2\pi RS^z_\mu\tilde\phi_{0\mu}\right)
\prod_{n>0}\frac{\left(a^\dagger_{n\mu}\right)^{m_{n\mu}}}{m_{n\mu}!}
|0\rangle,
\end{eqnarray}
where $|0\rangle$ is a vacuum ($a_{n\mu}|0\rangle=0$).
The constant $S^z_\mu$ is nothing but a quantum number of total
$S^z$ of each chain.
Since $l$ is an even integer, $S^z_\mu$ can take integer values only.
Therefore, in the limit $J_K\to\infty$, the ground state of the
total system is the state with $S^z_\mu=m_{n\mu}=0$ for $\mu=l$ and $r$.
The first excited states are fourfold degenerate and correspond to
$(S^z_l,S^z_r)=(\pm1,0),(0,\pm1)$ and $m_{n\mu}=0$.
The energy gap in this limit is then given by
\begin{equation}
E_g=\frac{\pi v}{l}2\pi R^2,
\label{E_g}
\end{equation}
which equals $\sqrt{3}\pi/2l$ at $\Delta=1/2$.
This gap value is shown as a dashed line in Fig.~\ref{fig:gap_d.5}.
It is clear that all the curves in Fig.~\ref{fig:gap_d.5} are
gradually approaching the dashed line as $L$ increases.
How the curves finally approach it in the $L\to\infty$ limit is
determined by the leading irrelevant operator $\widehat{O}_1$, whose
explicit form we may take
\begin{eqnarray}
\widehat{O}_1&\propto&
\cos\left\{
i2\pi R\left[\tilde\phi_r(0,0)-\tilde\phi_l(0,0)\right]
\right\}.
\label{O_1}
\end{eqnarray}
The correction to Eq.\ (\ref{E_g}) due to the operator $\widehat{O}_1$
can be obtained from lowest-order perturbation expansion.\cite{Cardy}
Since the degenerate first excited states $|S^z_l=1,S^z_r=0\rangle$
and $|S^z_l=0,S^z_r=1\rangle$ have a nonzero matrix element,
\begin{eqnarray}
&&\langle S^z_l=1,S^z_r=0|\widehat{O}_1|S^z_l=0,S^z_r=1\rangle
\nonumber\\
&&\propto
\langle0|\exp\left[-2\pi R\sum_{n=1}^l\frac{1}{\sqrt{\pi n}}
\left(a_{nr}-a^\dagger_{nr}-a_{nl}+a^\dagger_{nl}\right)
\right]|0\rangle
\nonumber\\
&&\propto
L^{-4\pi R^2},
\label{diff_+}
\end{eqnarray}
the degeneracy of these two states is lifted by an amount which
scales as $L^{-4\pi R^2}$.
The same is true for the other two degenerate states
$|S^z_l=-1,S^z_r=0\rangle$ and $|S^z_l=0,S^z_r=-1\rangle$.
On the other hand, the ground state energy does not change in first-order
perturbation. 
Hence we may expect that the leading correction to the energy gap
should be proportional to $L^{-4\pi R^2}$, which goes to zero faster than
the finite-size gap ($\propto L^{-1}$).
This $L$ dependence is indeed observed in our numerical data shown in
Fig.~\ref{fig:diff_gap_d.5}.
The data shows very clear power-law behavior with the exponent $4/3=4\pi
R^2$, in perfect agreement with the theory.
This can be regarded as a numerical proof of the presence of the leading
irrelevant operator with the scaling dimension $4\pi R^2$ at the
strong-coupling fixed point we discussed in Sec.~IIB.
We note that the energy gap $E^{(0)}_g$ used in
Fig.~\ref{fig:diff_gap_d.5} is the one at $J_K=\infty$, or equivalently,
the finite-size gap of an XXZ spin chain containing $l$ spins under the
open boundary condition. 
The reason why we have used $E^{(0)}_g$ rather than Eq.\ (\ref{E_g}) is to
reduce the effect of a bulk irrelevant operator, $\cos(2\phi/R)$, of
dimension $1/\pi R^2=3$.

\noindent

\begin{figure}
\narrowtext
\begin{center}
\leavevmode\epsfxsize=80mm
\epsfbox{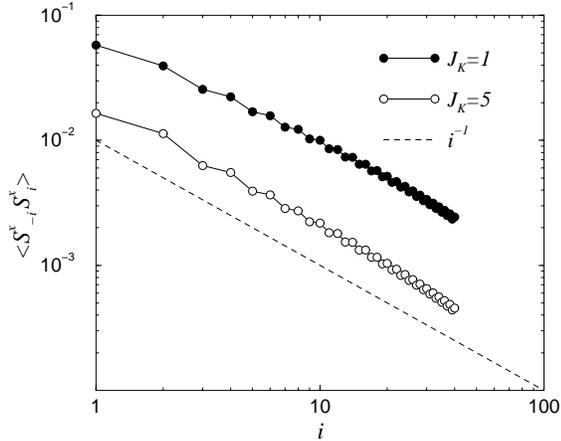}
\end{center}
\caption{
Correlation between $S^x_{-i}$ and $S^x_i$ calculated for $\Delta=0.5$
and $L=201$.
The dashed line corresponds to the $1/i$ decay obtained from the
perturbative calculation.
} 
\label{fig:S-iSi_x_d.5}
\end{figure}

\begin{figure}
\narrowtext
\begin{center}
\leavevmode\epsfxsize=80mm
\epsfbox{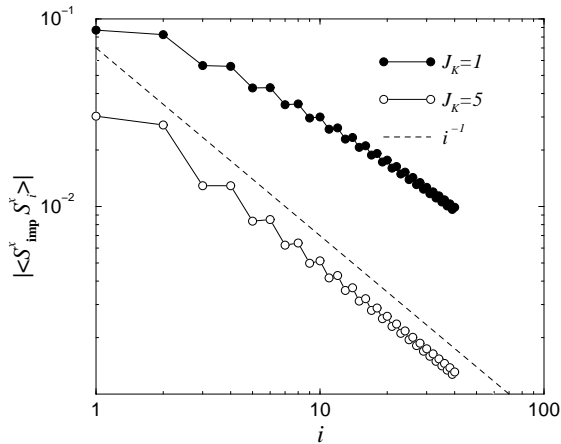}
\end{center}
\caption{
Correlation between $S^x_{\rm imp}$ and $S^x_i$ calculated for
$\Delta=0.5$ and $L=201$.
The dashed line represents the expected $i^{-1}$ behavior.
} 
\label{fig:SimpSi_x_d.5}
\end{figure}

Using the DMRG method, we have also calculated an equal-time two-point
spin correlation function, $\langle S_i^x S_j^x\rangle$, in the ground
state for $L=201$ ($S^z_{\rm tot}=0$).
According to our picture of the strong-coupling fixed point, the host
XXZ spin chain is effectively cut by a singlet in the low-energy limit
(Fig.~\ref{fig:rgflow+}). 
We naturally expect that correlations across the singlet should be much
weaker than correlations within one of the decoupled chains.
Our numerical results shown in Figs.~\ref{fig:S-iSi_x_d.5} and
\ref{fig:SimpSi_x_d.5} support this idea:
A correlation function across the singlet show power-law dependence on
$i$ with an exponent larger than that for a pure XXZ chain ($J_K=0$),
$2\pi R^2$.

The exponents for $\langle S^x_{-i}S^x_i\rangle$ and
$\langle S^x_{\rm imp}S^x_i\rangle$ can be obtained from the following
argument.
First we consider $\langle S^+_{-i}S^-_i\rangle$, which is equivalent to
$\langle S^x_{-i}S^x_i\rangle$.
Since it vanishes when the XXZ chain is completely decoupled, the nonzero
contribution is due to the leading irrelevant operator
$S^+_1S^-_{-1}+S^-_1S^+_{-1}$.
To first order in $\widehat{O}_1$ the correlator is
\begin{equation}
\langle S^+_{-i}S^-_i\rangle\propto
\int dt
\langle S^-_{-1}(t)S^+_{-i}(0)\rangle_l
\langle S^+_1(t)S^-_i(0)\rangle_r,
\label{S+S-}
\end{equation}
where the averages $\langle\quad\rangle_l$ and $\langle\quad\rangle_r$ are
evaluated for the ground state of each decoupled chain.
Since the scaling dimension of the boundary operators $S^\pm_{\mp1}$ is
$2\pi R^2$ and that of $S^\mp_{\pm i}$ is $\pi R^2$, we expect the
correlator to scale as
\begin{equation}
\langle S^+_{-i}S^-_i\rangle\propto i^{-6\pi R^2+1},
\label{S+S-_2}
\end{equation}
from which we get $\langle S^x_{-i}S^x_i\rangle\propto1/i$ for
$\Delta=1/2$.
The results in Fig.~\ref{fig:S-iSi_x_d.5} are consistent with this
perturbative calculation.

\noindent

\noindent

The correlation between $S^x_{\rm imp}$ and $S^x_i$ can be calculated
using the $1/J_K$ expansion, which can be justified in the low-energy
limit.
At $J_K=\infty$ the ground state of the whole system is a
direct product of $|S\rangle$, which is the singlet wave function of
$\mbox{\boldmath$S$}_{\rm imp}$ and $\mbox{\boldmath$S$}_0$, and the
ground states of the left and right decoupled spin chains, which we denote
$|l\rangle$ and $|r\rangle$.
We calculate correlation  function $\langle S^+_{\rm imp}S^-_i\rangle$ to
lowest order in the coupling between $\mbox{\boldmath$S$}_0$ and its
neighboring spin, $S^-_0S^+_1$:
\begin{eqnarray}
\langle S^+_{\rm imp}S^-_i\rangle&\sim&
\frac{1}{J_K}\langle S|S^+_{\rm imp}|T\rangle\langle T|S^-_0\rangle
\langle r|S^+_1S^-_i|r\rangle
\nonumber\\
&\propto&
(-1)^i i^{-3\pi R^2},
\label{Simp+Si-}
\end{eqnarray}
where $|T\rangle$ is a triplet state of $\mbox{\boldmath$S$}_{\rm imp}$
and $\mbox{\boldmath$S$}_0$ having excitation energy of order $J_K$.
The exponent $3\pi R^2(=1)$ is a sum of the dimensions of $S^+_1$ and
$S^-_i$.
The data for $J_K=5$ in Fig.~\ref{fig:SimpSi_x_d.5} is in excellent
agreement with the above calculation, although the data for $J_K=1$
is curving, which we think is due to a crossover to the true scaling
regime.

\subsection{Numerical results for $\Delta=-0.5$}

Here we present the numerical results for negative $\Delta$.
Using the DMRG method, we have calculated finite-size gap and spin
correlation functions for $\Delta=-0.5$,
where $R=1/\sqrt{6\pi}$ and $v=3\sqrt{3}/8$.

\begin{figure}
\narrowtext
\begin{center}
\leavevmode\epsfxsize=80mm
\epsfbox{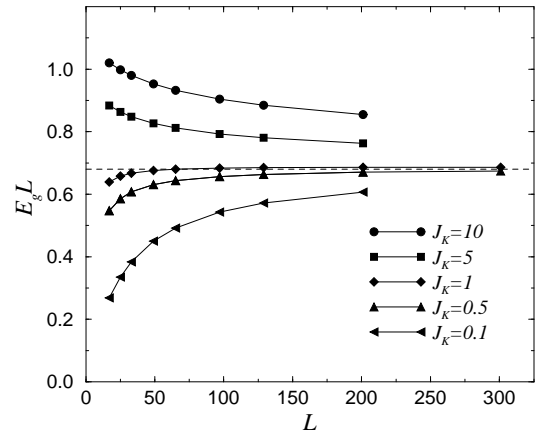}
\end{center}
\caption{Energy gap $E_g$ as a function of system size $L$ for
$\Delta=-0.5$.
The data are taken for $L=17$, 25, 33, 49, 65, 97, 129, 201, and 301.
The dashed line represents the infinite-$L$ limit,
$E_gL=\protect\sqrt{3}\pi/8$. 
}
\label{fig:gap_d-.5}
\end{figure}\noindent

Figure \ref{fig:gap_d-.5} shows the finite-size energy gap as a function
of the system size for $L=1$ (mod 4).
As in the last section, the gap is defined as difference between the
lowest energy in the sector $S^z_{\rm tot}=0$ and that in the sector
$S^z_{\rm tot}=1$.
We find that the normalized gap $E_gL$ increases for small $J_K$, while
it decreases for large $J_K$.
This is consistent with our picture of the renormalization flows
(Fig.~\ref{fig:rgflow-}).
For small $J_K$ the excitation gap is due to fluctuations of
$\mbox{\boldmath$S$}_{\rm imp}$ weakly coupled to the host spin chain.
This coupling is renormalized and becomes stronger as we saw in Sec.~IIA.
For large $J_K$, on the other hand, the host XXZ chain is almost cut by
a singlet, and the finite-size gap roughly corresponds to the
singlet-to-triplet excitation energy in half chains.
As $L$ increases, or equivalently, as the energy scale decreases, the
renormalized coupling between the almost decoupled chains becomes larger
(^^ ^^ healing''), leading to the decrease of the normalized
finite-size gap.
It is clear that all the curves in Fig.~\ref{fig:gap_d-.5} approach
the dashed line $E_gL=\sqrt{3}\pi/8=0.680\cdots$, which is the value one
expects for a single XXZ chain of length $L$.
Unlike in the case of $\Delta=0.5$, however, we have not been able to
obtain information on the scaling dimension of a leading irrelevant
operator from the numerical data.
A log-log plot of $|E_g-E_g^{(0)}|$ versus $L$ did not give straight
lines corresponding to power-law scaling.
This would mean that the systems we have studied ($L\sim200$) are
not large enough.

\begin{figure}
\narrowtext
\begin{center}
\leavevmode\epsfxsize=80mm
\epsfbox{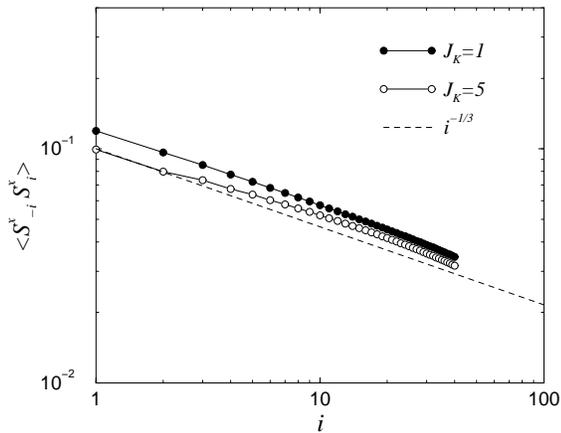}
\end{center}
\caption{
Correlation function $\langle S^x_{-i}S^x_i\rangle$ for $\Delta=-0.5$
and $L=201$.
The dashed line corresponds to the $i^{-1/3}$ decay.
}
\label{fig:S-iSi^x_d-.5}
\end{figure}\noindent

\begin{figure}
\narrowtext
\begin{center}
\leavevmode\epsfxsize=80mm
\epsfbox{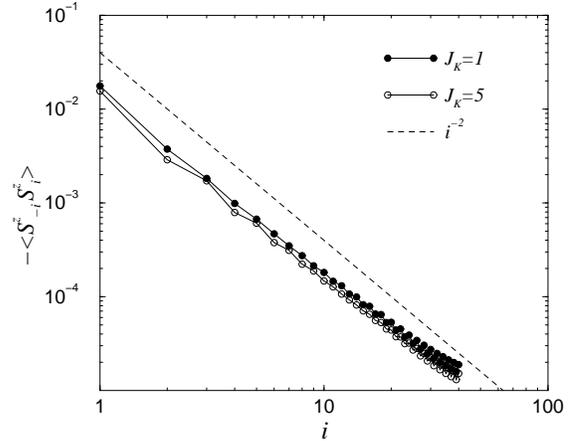}
\end{center}
\caption{
Correlation function $\langle S^z_{-i}S^z_i\rangle$ for $\Delta=-0.5$
and $L=201$.
The dashed line corresponds to the $i^{-2}$ behavior.
}
\label{fig:S-iSi^z_d-.5}
\end{figure}

Next we show the results of correlation functions which we computed
for the ground state of $L=201$ system ($S^z_{\rm tot}=0$)
Figure \ref{fig:S-iSi^x_d-.5} and \ref{fig:S-iSi^z_d-.5} show
correlation functions of $\mbox{\boldmath$S$}_{-i}$ and
$\mbox{\boldmath$S$}_i$.
The correlator $\langle S^x_{-i}S^x_i\rangle$ is positive and decays
like $i^{-1/3}$, while $\langle S^z_{-i}S^z_i\rangle$ is negative and
decays much faster like $i^{-2}$.
These features are exactly what we expect from our picture of the
low-energy fixed point (Figs.~\ref{fig:rgflow-} and \ref{fig:rgflow-2}).
Since the spin chain is well connected, the correlation functions
$\langle S^\alpha_{-i}S^\alpha_i\rangle$ should behave as in a pure
XXZ chain without an impurity spin.
That is, exponents of power-law decays should be the same as those in
the pure chain, although amplitudes of the correlators will depend on
$J_K$.
From Eqs.~(\ref{S^z}) and (\ref{S^+}) we see that at long distance
$S^z_i\sim d\phi/dx$ and $S^+_i\sim(-1)^ie^{i2\pi R\tilde\phi}$, whose
scaling dimensions are 1 and $\pi R^2=1/6$.
Hence $\langle S^z_{-i}S^z_i\rangle$ should decay as
$i^{-2}$ and $\langle S^x_{-i}S^x_i\rangle\propto i^{-1/3}$, in
agreement with the numerical result.\cite{error}

\noindent

\begin{figure}
\narrowtext
\begin{center}
\leavevmode\epsfxsize=80mm
\epsfbox{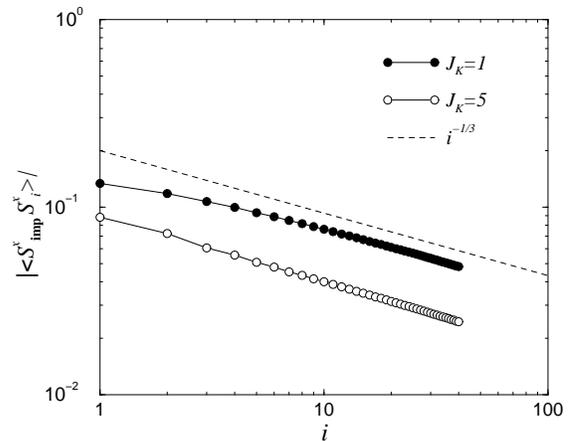}
\end{center}
\caption{
Correlation between $S^x_{\rm imp}$ and $S^x_i$ for $\Delta=-0.5$ and
$L=201$.
The dashed line corresponds to the $i^{-1/3}$ behavior.
}
\label{fig:SimpSi-x_d-.5}
\end{figure}

\begin{figure}
\narrowtext
\begin{center}
\leavevmode\epsfxsize=80mm
\epsfbox{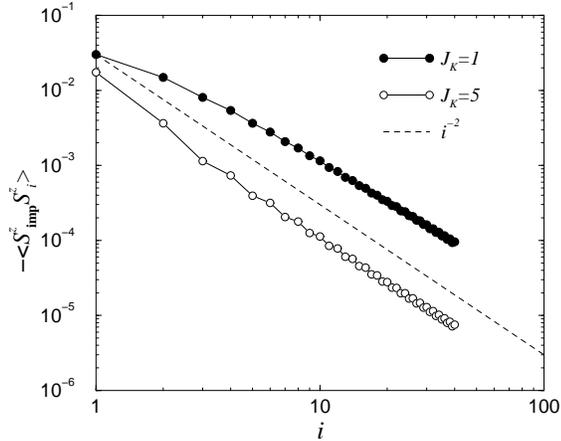}
\end{center}
\caption{
Correlation between $S^z_{\rm imp}$ and $S^z_i$ for $\Delta=-0.5$ and
$L=201$.
The dashed line corresponds to the expected $i^{-2}$ decay.
}
\label{fig:SimpSi-z_d-.5}
\end{figure}

We next discuss correlations between $\mbox{\boldmath$S$}_{\rm imp}$
and $\mbox{\boldmath$S$}_i$.
Since there is always a short-distance correlation between
$\mbox{\boldmath$S$}_{\rm imp}$ and
$\mbox{\boldmath$S$}_1+\mbox{\boldmath$S$}_{-1}$, we expect
$\langle S^\alpha_{\rm imp}S^\alpha_i\rangle
 \propto\langle(S^\alpha_1+S^\alpha_{-1})S^\alpha_i\rangle$ with a
smaller constant of proportion for larger $J_K$.
Noting that $S^x_1+S^x_{-1}$ corresponds to the staggered component in
a pure XXZ chain without the spin rotation of $\mbox{\boldmath$S$}_i$
($i>0$), we conclude that
$\langle S^x_{\rm imp}S^x_i\rangle\propto i^{-1/3}$ and
$\langle S^z_{\rm imp}S^z_i\rangle\propto i^{-2}$ for large $i$.
Our numerical results shown in Figs.~\ref{fig:SimpSi-x_d-.5} and
\ref{fig:SimpSi-z_d-.5} show exactly the feature discussed above.
Hence we conclude that the numerical results support our picture of
the low-energy fixed point.

\noindent

\noindent

\section{Conclusions}

In this paper we have studied the Kondo effect due to an extra spin
coupled to a gapless XXZ spin chain.
In our model the backward spinflip scattering is always a relevant
perturbation. 
At low energy the impurity spin is screened by a spin
in the host chain, and the characteristic energy scale, the Kondo
temperature $T_K$, has a power-law dependence on the Kondo coupling.
From the perturbative RG analysis for various limits, we have deduced
properties of strong-coupling, low-energy fixed points.
In the antiferromagnetic side ($0<\Delta\le1$) the host XXZ chain is cut
by the singlet into two separate chains.
On the other hand, in the ferromagnetic side ($-1<\Delta<0$) the singlet
does not harm the host spin chain in the low-energy limit.
This may be understood qualitatively by mapping the problem to a
spinless fermionic system using the Jordan-Wigner transformation.
The fermions have mutual repulsive (attractive) interactions in the
antiferromagnetic (ferromagnetic) region.
The singlet may then be viewed as an impurity potential for the fermions,
which can be a relevant or irrelevant perturbation, depending on the sign
of the mutual interactions.
Employing the known result for the spinless fermion system,\cite{Kane}
we can argue that the host spin chain is cut into two pieces in the
antiferromagnetic case whereas in the other case the singlet does not
affect the low-energy properties of the spin chain.

We have used the powerful DMRG method to numerically compute finite-size
energy gaps and correlation functions.
The numerical results are consistent with the RG analysis.
For $\Delta=0.5$, the normalized gap approaches the value for a
chain of half length, and the correlation function across
$\mbox{\boldmath$S$}_0$ decays much faster than in a pure spin
chain ($J_K=0$).
These results are explained successfully based on the RG analysis of the
strong-coupling fixed point (Fig.~\ref{fig:rgflow+}).
For $\Delta=-0.5$ we have found that the normalized gap approaches the
value for a spin chain without the Kondo impurity.
The correlation functions also show the same power-law
behavior as in the pure spin chain.
These results are consistent with our picture of the fixed point where
the host spin chain remains as a single chain through healing of a
coupling weakened by the singlet formation (Figs.~\ref{fig:rgflow-}
and \ref{fig:rgflow-2}).

\acknowledgements
A.F.~thanks N.\ Kawakami, N.\ Nagaosa, and N.V.\ Prokof'ev for useful
discussions on various aspects of the Kondo effect.
Numerical calculations were performed in part on NEC SX4 at the
Yukawa Institute for Theoretical Physics, Kyoto University.
This work was initiated when T.H.~stayed at the Yukawa Institute as
an ^^ ^^ Atom'' researcher.

\end{multicols}


\begin{references}
\bibitem{Nozieres} This is similar to the Nozi\`eres's discussion on
  effective interactions in the conventional Kondo
  problem.\cite{Nozieres2}
  The $1/J$ expansion can generate all the possible effective
  interactions, whose coefficients, however, depend on RG trajectories
  and therefore may have different sign and magnitude when starting
  the RG from the weak-coupling region.
\bibitem{reality} In reality, of course, the screening cloud is extended
over several ($\sim1/T_K$) sites.
\bibitem{locally} We can obtain a generic leading temperature
  dependence of $\delta\chi$ by applying a magnetic field to
  $\mbox{\boldmath$S$}_{\rm imp}$ only, although the standard
  definition of $\delta\chi$ is difference of the susceptibility of
  the whole system with and without an impurity spin.
\bibitem{error} In fig.~\ref{fig:S-iSi^z_d-.5} the data is slightly
  deviating from the $i^{-2}$ behavior for $i>30$.  However, the
  correlation of order $10^{-5}$ is already close to the estimated
  size of numerical error, and we do not think that the deviation
  is a serious problem.
\end{references}
\end{document}